\documentclass[aps,twocolumn,amsmath,amssymb,floatfix,
superscriptaddress]{revtex4}

\usepackage[dvips]{graphics}
\usepackage{color}
\definecolor{gold}{rgb}{0.85,0.66,0}
\definecolor{dred}{rgb}{0.6,0,0}

\begin{document}

\title{Multi-terminal magneto-transport in an interacting fractal 
network: a mean field study}

\author{Santanu K. Maiti}

\email{santanu@post.tau.ac.il}

\affiliation{School of Chemistry, Tel Aviv University, Ramat-Aviv,
Tel Aviv-69978, Israel}

\author{Arunava Chakrabarti}

\affiliation{Department of Physics, University of Kalyani, Kalyani,
West Bengal-741 235, India}

\begin{abstract}

Magneto-transport of interacting electrons in a Sierpinski gasket fractal 
is studied  within a mean field approach. We work out the three-terminal 
transport and study the interplay of the magnetic flux threading the planar 
gasket and the dephasing effect introduced by the third lead. For 
completeness we also provide results of the two-terminal transport in 
presence of electron-electron interaction. It is observed that dephasing 
definitely reduces the transport, while the magnetic field generates a  
continuum in the transmission spectrum signaling a band of extended 
eigenstates in this non-translationally invariant fractal structure.
The Hubbard interaction and the dephasing introduced by the third lead 
play their parts in reducing the average transmission, and opens up gaps 
in the spectrum, but can not destroy the continuum in the spectrum.

\end{abstract}

\maketitle

\section{Introduction}

Electronic transport in low dimensional systems is a key to explore 
important interference effects, local currents, switching mechanisms 
and other unique properties that are pre-requisites for estimating the 
potentiality of these systems as quantum interference 
devices~\cite{aviram,joachim,reed}. A considerable amount of work in 
this direction has already revealed the unique features of phase coherent 
electron transport through quantum dots (QD), Aharonov-Bohm (AB) rings, 
and model molecular systems~\cite{seki,storm,sigri,song,entin,aharon,
joe,jaya,ritter,saha,baran,buttiker,sant1,cook,cardamone,sant4}.

A major part of the studies in low-dimensions so far describes electron 
transport in the so called two-terminal nano- or mesoscopic 
devices~\cite{datta}. Indeed, this is a fast developing field, and has 
stimulated lot of theoretical work based on the non-equilibrium Green's 
function approach within the density functional 
theory~\cite{faleev,pala,xue,lambert}. 

Comparatively speaking, much smaller volume of literature on the three 
or four terminal electronic transport have come up in recent 
times~\cite{jaya,ritter,saha,baran,buttiker,sant1,cook,cardamone}. A 
two-terminal device is essentially a single path device. The transport 
here is marked by a sharp jump in the transmission phase, and is 
constrained by the Onsager relations of time reversal 
symmetry~\cite{onsager} and the current conservation~\cite{joe}. So, 
the inclusion of a third terminal that allows the current to flow out 
of the system and breaks the unitarity condition, is likely to be useful 
in extracting useful information related to the quantum coherence is 
low-dimensional systems~\cite{buttiker2,buttiker3}.

\begin{figure}[ht]
{\centering \resizebox*{8cm}{4.5cm}{\includegraphics{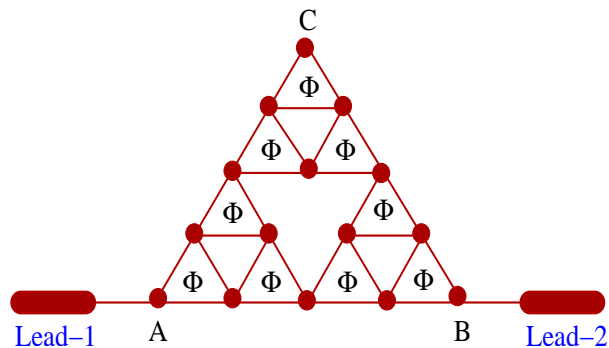}}\par}
\caption{(Color online). A $3$-rd generation Sierpinski gasket is attached 
to two semi-infinite $1$D metallic leads, viz, lead-1 and lead-2. Each 
elementary plaquette of the gasket is penetrated by a magnetic flux $\phi$. 
The filled red circles correspond to the positions of the atomic sites.}
\label{spg1}
\end{figure}

\begin{figure}[ht]
{\centering \resizebox*{8cm}{6cm}{\includegraphics{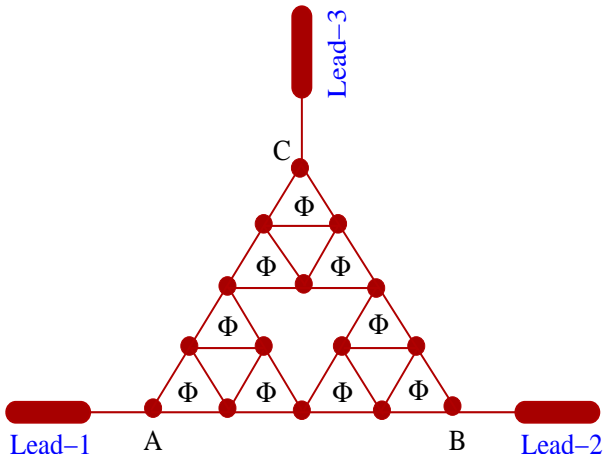}}\par}
\caption{(Color online). A $3$-rd generation Sierpinski gasket is attached 
to three semi-infinite $1$D metallic leads, viz, lead-1, lead-2 and lead-3. 
Each elementary plaquette of the gasket is penetrated by a magnetic flux 
$\phi$. The filled red circles correspond to the positions of the atomic 
sites.}
\label{spg2}
\end{figure}

In this communication we undertake an in-depth study of the three-terminal 
magneto-transport of interacting electrons in a fractal network. 
Specifically, we choose a fractal geometry following a Sierpinski gasket 
(SPG)~\cite{domany,rammal,banavar,ghez,schwalm1,andrade1,schwalm2,andrade2,
schwalm3}. Such a planar gasket is shown in Figs.~\ref{spg1} and \ref{spg2}, 
and can be thought to be equivalent to a self-similar arrangement of single 
level QD's~\cite{kubala1,kubala2} sitting at the vertices of each elementary 
triangle. With the present day advancement in lithography 
practically any design can be tailor made, and the present work thus offers 
an excellent opportunity to study the simultaneous effects of magnetic 
field, electron-electron interaction and the dephasing caused by the 
introduction of a third lead in the system.

Our motivation behind the present work is two-fold. First, we observe 
that the multi-terminal transport in systems with multiply connected 
geometry is a very little addressed (or, unaddressed) problem. In 
particular, the in-built self similarity of systems such as the SPG opens 
up the possibility of investigating the tunneling or switching aspect of 
these systems at arbitrarily small scales of energy. Secondly, it is 
essential, for a completeness in the understanding of the spectral 
properties of fractal networks to know, if the well known multi-fractal, 
Cantor set energy spectrum of non-interacting electrons~\cite{domany,rammal,
banavar,ghez,schwalm1,andrade1,schwalm2,andrade2,schwalm3} still retains 
its character even in the presence of electron-electron interaction or 
dephasing caused by a third electrode. In a recent work, the effect of 
electron-electron interactions on the persistent current in a closed loop 
SPG has been reported~\cite{sant2}, but no result exists for open 
self-similar systems connected to electron reservoirs by multiple leads.

On its own merit, the effect of electron-electron interaction on the 
spectral properties is of great importance. Several experiments done on 
fractal networks have studied the magnetoresistance, the
superconductor-normal phase boundaries on Sierpinski gasket wire
networks~\cite{gordon1,gordon2,gordon3,korshu,meyer}. These experiments 
pioneered the actual observational studies of spectral properties and 
flux quantization effects on planar networks and the Aharonov-Bohm effect 
in systems without translational invariance. Although in an early paper 
the problem of interacting electrons on a percolating cluster that displays 
a fractal geometry~\cite{nedellec}, has been addressed, to the best of our 
knowledge, no rigorous effort has been made so far to unravel the effect 
of an interplay of electron-electron interaction and an external magnetic 
field on deterministic networks such as a Sierpinski gasket (SPG), even 
at a mean field level.

Thus, apart from a critical investigation of the possibility of a fractal 
{\em device}, the present work is also likely to throw light on the 
fundamental spectral properties of the {\em deterministically disordered} 
systems.

We find quite interesting results. Working within a tight-binding framework 
we develop a mean field method of studying the three-terminal transport 
in interacting systems. The method is then applied to a planar SPG and we 
show that, dephasing definitely reduces the corner-to-corner propagation 
of electrons, with or without the electron-electron interaction. The 
magnetic field, in the absence of the electron-electron interaction 
generates an apparent continuum in the transmission spectrum. The 
{\em continuum} has already been observed in several fractal 
lattices~\cite{arun2,arun3,schwalm4} including an SPG~\cite{arun1}, and 
is practically unaffected even when the electron-electron interaction 
is switched `on'. A detailed study on the effect of positioning of the 
third electrode is also made, and the transport properties of an anisotropic 
gasket are compared with its isotropic counterpart. For a comparative study, 
the results of the two-terminal transport are also presented along with the 
three-terminal cases.

In what follows, we present our model quantum system in Section II. The 
essentials of the mean field calculation are discussed in Section III. 
Section IV contains the numerical results and related discussions, and we 
draw our conclusions in Section V.	 
	
\section{The model quantum systems}

Let us begin by referring to Fig.~\ref{spg1} where a $3$-rd generation SPG 
is attached to two semi-infinite one-dimensional ($1$D) metallic leads, 
namely, lead-1 and lead-2 via the atomic sites A and B. Each elementary 
plaquette of the gasket is threaded by a magnetic flux $\phi$ (measured 
in unit of the elementary flux quantum $\phi_0=ch/e$). The filled red 
circles correspond to the positions of atomic sites in the SPG. In a 
Wannier basis the tight-binding Hamiltonian of the interacting gasket 
reads,
\begin{eqnarray}
H_{\mbox{SPG}} & = &\sum_{i,\sigma}\epsilon_{i\sigma} c_{i\sigma}^{\dagger} 
c_{i\sigma} + \sum_{\langle ij \rangle,\sigma} v \left[e^{i\theta} 
c_{i\sigma}^{\dagger} c_{j\sigma} + h.c. \right] \nonumber \\
&  & +\sum_i U c_{i\uparrow}^{\dagger}c_{i\uparrow} c_{i\downarrow}^{\dagger}
c_{i\downarrow}
\label{spgham}
\end{eqnarray}
where, $\epsilon_{i\sigma}$ is the on-site energy of an electron at
the site $i$ of spin $\sigma$ ($\uparrow,\downarrow$) and $v$ is the
nearest-neighbor hopping strength. In the case of an anisotropic SPG,
the anisotropy is introduced only in the nearest-neighbor hopping
integral $v$ which takes on values $v_x$ and $v_y$ for hopping along
the {\it horizontal} and the {\it angular} bonds, respectively. Due to
the presence of magnetic flux $\phi$, a phase factor $\theta=2\pi\phi/3$
appears in the Hamiltonian when an electron hops from one site to another
site. A negative sign comes in when the electron hops in the reverse 
direction. As the magnetic filed associated with $\phi$ does not penetrate 
any part of the circumference of the elementary triangle, we ignore the 
Zeeman term in the above tight-binding Hamiltonian (Eq.~\ref{spgham}). 
$c_{i\sigma}^{\dagger}$ and $c_{i\sigma}$ are the creation and annihilation 
operators, respectively, of an electron at the site $i$ with spin $\sigma$. 
$U$ is the strength of on-site Coulomb interaction. The Hamiltonian for 
the non-interacting leads can be expressed as,
\begin{equation}
H_{\mbox{lead}} = \sum_i \epsilon_0 c_i^{\dagger} c_i + \sum_{<ij>} 
t_0 \left(c_i^{\dagger} c_j + c_j^{\dagger} c_i \right)
\label{leadham}
\end{equation}
where different parameters correspond to their usual meaning. These leads
are directly coupled to the gasket where the hopping integral between the
lead-1 and gasket is $\tau_1$, and, it is $\tau_2$ between the gasket and
lead-2. With this setup we investigate two-terminal electron transport
through an SPG.

For three-terminal quantum transport we connect an additional lead with
the gasket. A schematic view of a $3$-rd generation SPG attached to three
semi-infinite $1$D metallic leads, viz, lead-1, lead-2 and lead-3 is shown
in Fig.~\ref{spg2}. The gasket and the side-attached leads are described
by the same prescriptions as described above.

\section{The mean field approach}

\subsection{Decoupling of the interacting Hamiltonian}

Before going to the calculation of electronic transmission probability
through the interacting model of an SPG described by the tight-binding 
Hamiltonian given in Eq.~\ref{spgham}, first we decouple the interacting 
Hamiltonian using the generalized Hartree-Fock approach~\cite{kato,kam,
sant5}. The full Hamiltonian is completely decoupled into two parts. One 
is associated with the up-spin electrons, while the other is with the 
down-spin electrons. The on-site potentials get modified appropriately, 
and are given by,
\begin{equation}
\epsilon_{i\uparrow}^{\prime}=\epsilon_{i\uparrow} + U \langle 
n_{i\downarrow} \rangle
\label{equ200}
\end{equation}
\begin{equation}
\epsilon_{i\downarrow}^{\prime}=\epsilon_{i\downarrow} + U \langle 
n_{i\uparrow} \rangle
\label{equ300}
\end{equation}
where, $n_{i\sigma}=c_{i\sigma}^{\dagger} c_{i\sigma}$ is the number 
operator. With these site energies, the full Hamiltonian (Eq.~\ref{spgham})
can be written in the decoupled form (in the mean field (MF) approximation) 
as,
\begin{eqnarray}
H_{\mbox{MF}} &=&\sum_i \epsilon_{i\uparrow}^{\prime} n_{i\uparrow} + 
\sum_{\langle ij \rangle} v \left[e^{i\theta} c_{i\uparrow}^{\dagger} 
c_{j\uparrow} + e^{-i\theta} c_{j\uparrow}^{\dagger} 
c_{i\uparrow}\right] \nonumber \\
& + & \sum_i \epsilon_{i\downarrow}^{\prime} n_{i\downarrow} + \sum_{\langle 
ij \rangle} v \left[e^{i\theta} c_{i\downarrow}^{\dagger} c_{j\downarrow}
+ e^{-i\theta} c_{j\downarrow}^{\dagger} c_{i\downarrow}\right] \nonumber \\
& - & \sum_i U \langle n_{i\uparrow} \rangle \langle n_{i\downarrow} 
\rangle \nonumber \\
&=& H_{\mbox{SPG},\uparrow} + H_{\mbox{SPG},\downarrow} - 
\sum_i U \langle n_{i\uparrow} \rangle \langle n_{i\downarrow} \rangle
\label{equ400} 
\end{eqnarray}
where, $H_{\mbox{SPG}, \uparrow}$ and $H_{\mbox{SPG},\downarrow}$ 
correspond to the effective tight-binding Hamiltonians for the up and 
down spin electrons, respectively. The last term is a constant term 
which provides a shift in the total energy.

\subsection{Self consistent procedure}

With these decoupled Hamiltonians ($H_{\mbox{SPG},\uparrow}$ and 
$H_{\mbox{SPG},\downarrow}$) of up and down spin electrons, now we 
start our self consistent procedure considering initial guess values 
of $\langle n_{i\uparrow} \rangle$ and $\langle n_{i\downarrow} \rangle$. 
For these initial set of values of $\langle n_{i\uparrow} \rangle$ and 
$\langle n_{i\downarrow} \rangle$, we numerically diagonalize the up and 
down spin Hamiltonians. Then we calculate a new set of values of 
$\langle n_{i\uparrow} \rangle$ and $\langle n_{i\downarrow} \rangle$. 
These steps are repeated until a self consistent solution is achieved.

\subsection{Two-terminal quantum system}

Now we are at the stage of calculating electron conduction across an SPG.

To determine two-terminal conductance ($g$) of the gasket, we use Landauer 
conductance formula~\cite{datta}. At much low temperature and bias voltage 
it can be expressed as,
\begin{equation}
g=\frac{e^2}{h} \left(T_{\uparrow} + T_{\downarrow}
\right)
\label{equ1}
\end{equation}
where, $T_{\uparrow}$ and $T_{\downarrow}$ correspond to the transmission 
probabilities of up and down spin electrons, respectively, across the SPG.
Since no spin-flip scattering term exists in the Hamiltonian 
(Eq.~\ref{spgham}), spin-flip transmission probabilities will not 
appear in Eq.~\ref{equ1}. In terms of the Green's function of the gasket 
and its coupling to side-attached leads, transmission probability can be 
written in the form~\cite{datta},
\begin{equation}
T_{\sigma}={\mbox{Tr}} \left[\Gamma_1 \, G_{\mbox{SPG},\sigma}^r \, 
\Gamma_2 \, G_{\mbox{SPG},\sigma}^a\right]
\label{equ2}
\end{equation}
where, $\Gamma_1$ and $\Gamma_2$ describe the coupling of the SPG to the 
lead-$1$ and lead-$2$, respectively. Here, $G_{\mbox{SPG},\sigma}^r$ and 
$G_{\mbox{SPG},\sigma}^a$ are the retarded and advanced Green's functions, 
respectively, of the SPG including the effects of the leads. Now, for 
the complete system i.e., the SPG and two leads, the Green's function 
is expressed as,
\begin{equation}
G_{\sigma}=\left(E-H_{\sigma}\right)^{-1}
\label{equ3}
\end{equation}
where, $E$ is the energy of the injecting electron. Evaluation of this
Green's function needs the inversion of an infinite matrix, which is
really a difficult task, since the full system consists of the finite
size gasket and two semi-infinite $1$D leads. However, the full system 
can be partitioned into sub-matrices corresponding to the individual 
sub-systems and the Green's function for the gasket can be effectively 
written as, 
\begin{equation}
G_{\mbox{SPG},\sigma}=\left(E-H_{\mbox{SPG},\sigma}-\Sigma_1-\Sigma_2 
\right)^{-1}
\label{equ4}
\end{equation}
where, $\Sigma_1$ and $\Sigma_2$ are the self-energies due to coupling 
of the gasket to the lead-$1$ and lead-$2$, respectively. All information 
of the coupling are included into these self-energies.

\subsection{Three-terminal quantum system}

In order to calculate the conductance in three-terminal SPG, we use 
B\"{u}ttiker formalism, an elegant and simple way to study electron 
transport through multi-terminal mesoscopic systems. In this formalism
we treat all the leads (current and voltage leads) on an equal footing 
and extend the two-terminal linear response formula to get the conductance 
between the terminals, indexed by $p$ and $q$, in the 
form~\cite{datta,buttiker2},
\begin{equation}
g_{pq}=\frac{e^2}{h} \left(T_{pq,\uparrow} + T_{pq,\downarrow}\right)
\label{equ5}
\end{equation}
where, $T_{pq,\sigma}$ gives the transmission probability of an electron
with spin $\sigma$ ($\uparrow,\downarrow$) from the lead-$p$ to lead-$q$.
\begin{figure}[ht] 
{\centering \resizebox*{8.5cm}{9cm}
{\includegraphics{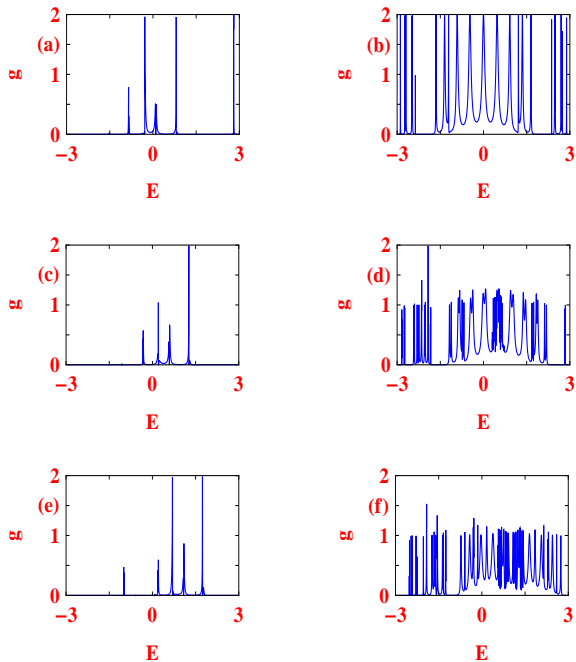}}\par}
\caption{(Color online). Two-terminal conductance $g$ as a function
of energy $E$ for a $4$-th generation isotropic ($v_x=v_y=1$) Sierpinski 
gasket ($N=42$), where lead-1 and lead-2 are connected at the positions
A and B, respectively. The first and second columns correspond to $\phi=0$
and $0.25$, respectively, where the upper, middle and lower panels 
represent $U=0$, $1$ and $2$, respectively.}
\label{twotermsym}
\end{figure}
Now, similar to Eq.~\ref{equ2} the transmission probability $T_{pq,\sigma}$
can be expressed in terms of the SPG-lead coupling matrices and the 
effective Green's function of the SPG as~\cite{datta},
\begin{equation}
T_{pq,\sigma}={\mbox{Tr}} \left[\Gamma_p \, G_{\mbox{SPG},\sigma}^r \, 
\Gamma_q \, G_{\mbox{SPG},\sigma}^a\right].
\label{equ6}
\end{equation}
In the presence of multi-leads, the effective Green's function of the 
SPG becomes (extension of Eq.~\ref{equ4})~\cite{datta},
\begin{equation}
G_{\mbox{SPG},\sigma}=\left(E-H_{\mbox{SPG},\sigma}-\sum_p \Sigma_p 
\right)^{-1}
\label{equ7}
\end{equation}
where, $\Sigma_p$ is the self-energy due to coupling of the SPG to the 
lead-$p$ and the sum over $p$ runs from $1$ to $3$.

In the present work we inspect all the essential features of magnetic
response of an SPG network at absolute zero temperature and use the
units where $c=h=e=1$. Throughout our numerical work we set
$\epsilon_{i\uparrow}=\epsilon_{i\downarrow}=0$ for all $i$ and choose
the nearest-neighbor hopping strength $v=1$. In the anisotropic case
we select $v_x = 1$ and $v_y = 2$ throughout. For side-attached leads, 
\begin{figure}[ht] 
{\centering \resizebox*{8.5cm}{9cm}
{\includegraphics{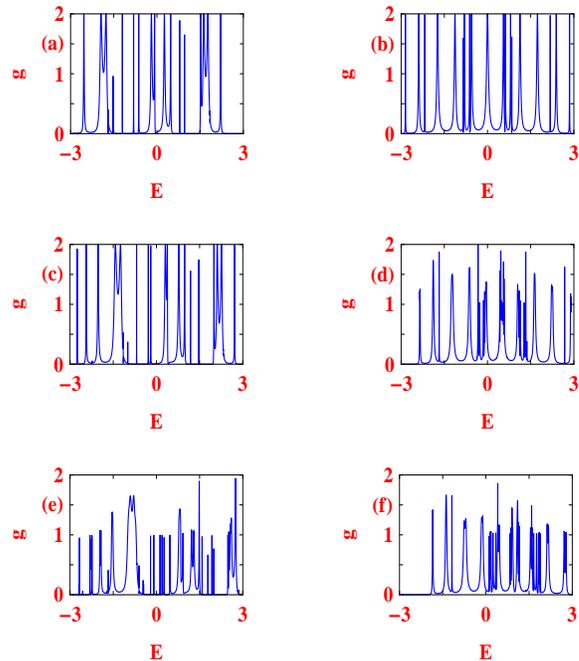}}\par}
\caption{(Color online). Two-terminal conductance $g$ as a function
of energy $E$ for a $4$-th generation anisotropic ($v_x=1$ and $v_y=2$) 
Sierpinski gasket ($N=42$), where lead-1 and lead-2 are connected at the 
positions A and B, respectively. The first and second columns correspond 
to $\phi=0$ and $0.25$, respectively, where the upper, middle and lower 
panels represent $U=0$, $1$ and $2$, respectively.}
\label{twotermasym}
\end{figure}
the on-site energy ($\epsilon_0$) and nearest-neighbor hopping strength
($t_0$) are fixed at $0$ and $1.5$, respectively. The hopping integrals
($\tau_1$, $\tau_2$ and $\tau_3$) between the leads and SPG are set at
$0.5$.  Energy scale is measured in unit of $v$. All the essential
features of electron transport are obtained both for an isotropic gasket 
and its anisotropic counterpart.

\section{Numerical results and discussion}

\subsection{Two-terminal quantum transport}

The two-terminal conductance of a $4$-th generation isotropic gasket is 
shown in Fig.~\ref{twotermsym} with the leads connected in the positions 
A and B only. The left panel shows the conductance in zero magnetic field, 
and represents the familiar fragmented, scanty distribution of the 
transmission resonances that mark such a lattice. The on-site Hubbard 
interaction displaces the resonance peaks, but no marked changes in the 
spectrum is observed.

With the magnetic field turned `on' there is however, a remarkable change. 
The spectrum apparently exhibits continuous parts, which have previously 
been reported~\cite{arun1} to support extended single particle states
for spinless, non-interacting electrons. We observe here that, the 
electron-electron interaction, though reduces the overall conductance of 
the system, preserves the continuum at the central part of the spectrum. 
As the interaction $U$ is increased, the central continuum is more or less 
undisturbed, though the average transmission amplitude still exhibits lower 
values compared to its $U=0$ counterpart. In addition, there is a signature 
of opening up of new gaps in the spectrum as it appears in 
Fig.~\ref{twotermsym}(f).

In Fig.~\ref{twotermasym} we show the two-terminal conductance across an 
SPG of the same size as before, but now with a hopping anisotropy such that, 
$v_x = 1$ and $v_y = 2$. The effect of hopping anisotropy is, in general, to 
\begin{figure}[ht] 
{\centering \resizebox*{8.5cm}{7.5cm}
{\includegraphics{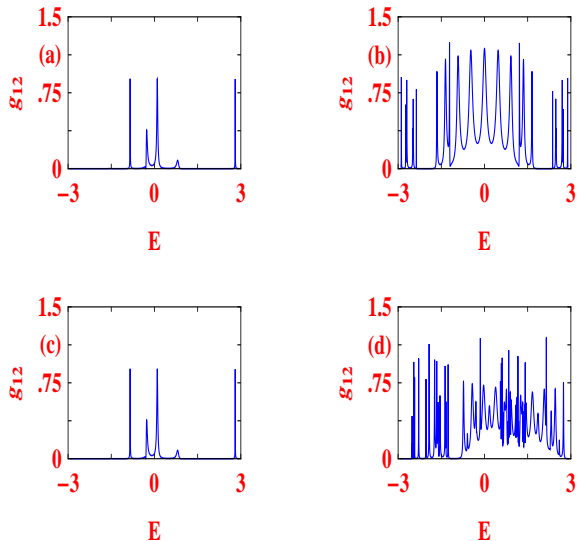}}\par}
\caption{(Color online). Three-terminal conductance $g_{12}$ as a function
of energy $E$ for a $4$-th generation isotropic ($v_x=v_y=1$) Sierpinski 
gasket ($N=42$), where lead-1, lead-2 and lead-3 are connected at the 
positions A, B and C, respectively. For this arrangement of side attached
leads (symmetric configuration), all $g_{ij}$'s ($i \neq j$) are identical
to each other. (a) $U=0$, $\phi=0$; (b) $U=0$, $\phi=0.25$; (c) $U=2$, 
$\phi=0$ and (d) $U=2$, $\phi=0.25$.}
\label{threetermsym1}
\end{figure}
increase the overall conductance. The sparse spectrum in the left panel of 
Fig.~\ref{twotermsym} is now replaced by a substantial density of resonant 
transmission peaks. The introduction of $U$ still reduces the average 
conductance, but the action is now somewhat `delayed'. We need a larger 
value of $U$  to generate a noticeable change in the conductance spectrum. 
This is evident from the left panel of Fig.~\ref{twotermasym}. On the right 
panel, the interplay of the magnetic field and the electron-electron 
interaction is displayed. The central {\em continuum} of the isotropic 
gasket case is still there, but with numerous conductance minima somewhat 
obscuring the continuity of this part. As $U$ increases, one observes 
genuine anti-resonances in the previously sustained continuum in the 
spectrum.

\subsection{Three-terminal quantum transport}

We now discuss the three-terminal transport in a SPG network. On the left 
panel of Fig.~\ref{threetermsym1} the results of the conductance between 
the leads $1$ and $2$ are shown. The third terminal $3$ connects to 
\begin{figure}[ht] 
{\centering \resizebox*{8.5cm}{9cm}{\includegraphics{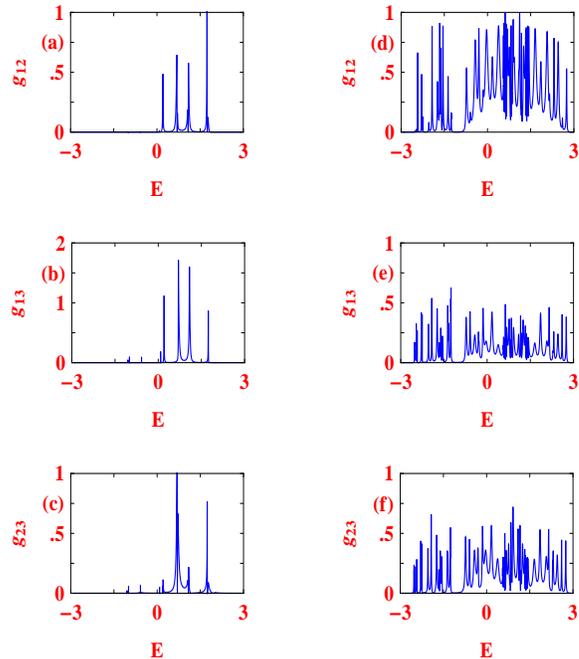}}\par}
\caption{(Color online). Three-terminal conductance $g_{ij}$ as a function
of energy $E$ for a $4$-th generation isotropic ($v_x=v_y=1$) Sierpinski 
gasket ($N=42$), where lead-1 and lead-2 are connected at the positions A 
and B, respectively. Lead-3 is attached at the centre of the line BC.
This is the so-called asymmetric configuration of the leads. First column
corresponds to $\phi=0$, while the second column represents $\phi=0.25$. 
For all these spectra we set $U=2$.}
\label{threetermasym}
\end{figure}
an electron reservoir. The effect of dephasing is obvious. The spectrum in 
zero magnetic field is even thinner now, and the interaction $U$ plays its 
part. The magnitude of the resonance peaks get suppressed compared to the 
coherent transport in the two-terminal case. A fine scrutiny (not shown 
here) will reveal a level broadening due a dominant phase randomizing 
effect~\cite{sant3} leading to a loss of phase coherence. Strangely enough, 
in presence of the magnetic field, the dephasing effect is just not enough 
to destroy or alter appreciably, the central continuous part of the 
conductance spectrum. We have carried out extensive numerical analysis of 
the continuum in the conductance spectrum, and it appears that the band 
remains intact over finer scales of energy interval and even for a larger 
sized SPG. Therefore, we are tempted to conclude that there exists a band 
of extended eigenstates (characterized by finite transmittivity) in an 
SPG fractal threaded by a magnetic field perpendicular to the plane of 
the gasket and that, its a quite {\em robust band}, unperturbed by either 
the electron-electron interaction, or the presence of any dephasing effect.

In the asymmetric case however, the conductance is strongly sensitive on 
the lead positions. For example, with $v_x=1$ and $v_y=2$, the conductance 
$g_{12}$ across the $1$-$2$ terminals is much larger compared to $g_{23}$ 
or $g_{13}$, particularly in presence of the magnetic field. The apparent 
continuum in the spectrum persists, as before, even with an increasing 
value of $U$. The results are displayed in Fig.~\ref{threetermasym}.

\section{Conclusion}

In conclusion, we have examined in details, the three-terminal transport 
in a planar Sierpinski gasket fractal with a magnetic field threading each 
elementary triangle. The electron-electron interaction is considered in the 
Hubbard form, and the Hamiltonian is solved within a generalized Hartree-Fock 
scheme. The sensitivity of the conductance spectrum on the positioning of the 
leads is studied extensively, and some prototype results have been displayed 
and discussed. It is seen that the major role of the third lead is to bring 
down the average conductance of the system. The same role is played by the 
Hubbard interaction. The anisotropic gasket has been found to be more 
conducting than its isotropic counterpart with or without the 
electron-electron interaction. However, the curious result is that, the 
continuum of states generated in the spectrum of an SPG as the magnetic 
field is turned on, is practically unaffected by the dephasing effect 
caused by the additional electrode or by the on-site Hubbard interaction.


\begin{thebibliography}{99}

\bibitem{aviram} A. Aviram and M. A. Ratner, Chem. Phys. Lett. \textbf{29}, 
277 (1974).

\bibitem{joachim} C. Joachim, J. K. Gimzewski, and A. Aviram, Nature 
\textbf{408}, 541 (2000).

\bibitem{reed} M. A. Reed, C. Zhou, C. J. Muller, T. P. burgin, and 
J. M. Tour, Science \textbf{278}, 252 (1997).


\bibitem{seki} T. Sekitani, U. Zschieschang, H. Klauk, and T. Someya, 
Nat. Mater. \textbf{9}, 1015 (20100.

\bibitem{storm} R. de Picciotto, H. L. Stormer, L. N. Pfeiffer, 
K. W. Baldwin, and K. W. West, Nature \textbf{411}, 51 (2001).

\bibitem{sigri} M. Sigrist, A. Fuhrer, T. Ihn, K. Ensslin, S. Ulloa, 
W. Wegescheider, and M. Bichler, Phys. Rev. Lett. \textbf{93}, 066802 (2004).

\bibitem{song} H. Song, Y. Kim, Y. H. Jang, H. Jeong, M. A. Reed, and
T. Lee, Nature \textbf{462}, 1039 (2009).


\bibitem{entin} O. Entin-Wohlman, A. Aharony, Y. Imry, Y. Levinson, 
and A. Schiller, Phys. Rev. Lett. \textbf{88}, 166801 (2002).

\bibitem{aharon} A. Aharony, O. Entin-Wohlman, B. I. Halperin, and 
Y. Imry, Phys. Rev. B \textbf{66}, 115311 (2002).

\bibitem{joe} Y. S. Joe, E. R. Hadin, and A. M. Satanin, Phys. Rev. B 
\textbf{76}, 085419 (2007).


\bibitem{jaya} T. Jayasekara and J. W. Mintmire, Nanotechnology \textbf{18}, 
424033 (2007).

\bibitem{ritter} C. Ritter, M. Pacheco, P. Orellana, and A. Latge, 
J. Appl. Phys. \textbf{106}, 104303 (2009).

\bibitem{saha} K. K. Saha, W. Lee, J. bernholc, and V. Meunier, J. Chem. 
Phys. \textbf{131}, 164105 (2009).

\bibitem{baran} H. U. Baranger, D. P. Di Vincenzo, R. A. Jalabert, and 
A. D. Stone, Phys. Rev. B \textbf{44}, 10637 (1991).

\bibitem{buttiker} M. B\"{u}ttiker, Phys. Rev. Lett. \textbf{57}, 1761 (1986).

\bibitem{sant1} S. K. Maiti, Solid State Commun. \textbf{150}, 1269 (2010).

\bibitem{cook} B. G. Cook, P. Dignard, and K. Varga, Phys. Rev. B 
\textbf{83}, 205105 (2011).

\bibitem{cardamone} D. M. Cardamone, C. A. Stafford, and S. Majumdar, 
Nano Lett. \textbf{6}, 2422 (2006).

\bibitem{sant4} P. Dutta, S. K. Maiti, and S. N. Karmakar, Org. Electron. 
\textbf{11}, 1120 (2010).

\bibitem{datta} S. Datta, {\em Electronic transport in mesoscopic systems},
Cambridge University Press, Cambridge (1997).


\bibitem{faleev} S. V. Faleev, F. L\'{e}onard, D. A. Stewart, and M. 
van Schilfgaarde, Phys. Rev. B \textbf{71}, 195422 (2005).

\bibitem{pala} J. J. Palacios, A. J. P\'{e}rez-Jim\'{e}nez, E. Louis, 
E. SanFabi\'{a}n, and J. A. Verg\'{e}s, Phys. Rev. Lett. \textbf{90}, 
106801 (2003).

\bibitem{xue} Y. Xue, S. Datta, and M. Ratner, J. Chem. Phys. \textbf{115}, 
4292 (2001).

\bibitem{lambert} S. Sanvito, C. J. Lambert, J. H. Jefferson, and 
A. M. Bratkovsky, Phys. Rev. B \textbf{59}, 11936 (1999).

\bibitem{onsager} L. Onsager, Phys. Rev. \textbf{38}, 2265 (1931).

\bibitem{buttiker2} M. B\"{u}ttiker, Phys. Rev. B \textbf{33}, 3020 (1986).

\bibitem{buttiker3} M. B\"{u}ttiker, IBM J. Res. Dev. \textbf{32}, 63 (1988).


\bibitem{domany} E. Domany, S. Alexander, D. Bensimon, and L. P. Kadanoff,
Phys. Rev. \textbf{28}, 3110 (1982).

\bibitem{rammal} R. Rammal and G. Toulose, Phys. Rev. Lett. \textbf{49},
1194 (1982).

\bibitem{banavar} J. R. Banavar, L. Kadanoff, and A. M. M. Pruisken,
Phys. Rev. B \textbf{31}, 1388 (1984).

\bibitem{ghez} J. M. Ghez, Y. Y. Wang, R. Rammal. B. Pannetier, and
J. B. Bellisard, Solid State Commun. \textbf{64}, 1291 (1987).

\bibitem{schwalm1} W. A. Schwalm and M. K. Schwalm, Phys. Rev. B
\textbf{39}, 12872 (1989).

\bibitem{andrade1} R. F. S. Andrade H. J. Schellnhuber, Europhys. Lett.
\textbf{10}, 73 (1989).

\bibitem{schwalm2} W. A. Schwalm and M. K. Schwalm, Phys. Rev. B
\textbf{44}, 382 (1991).

\bibitem{andrade2} R. F. S. Andrade and H. J Schellnhuber, Phys. Rev. B
\textbf{44}, 13213 (1991).

\bibitem{schwalm3} W. A. Schwalm and M. K. Schwalm, Phys. Rev. B
\textbf{47}, 7847 (1993).

\bibitem{kubala1} B. Kubala and J. K\"{o}nig, Phys. Rev. B \textbf{67}, 
205303 (2003); Phys. Rev. B \textbf{65}, 245301 (2002).

\bibitem{kubala2} B. Kubala and J. K\"{o}nig, Phys. Rev. B \textbf{65}, 
245301 (2002).

\bibitem{sant2} S. K. Maiti and A. Chakrabarti, Phys. Rev. B \textbf{82}, 
184201 (2010).

\bibitem{gordon1} J. M. Gordon, A. M. goldman, J. Maps, D. Costello,
R. Tiberio, and B. Whitehead, Phys. Rev. Lett. \textbf{56}, 2280 (1986).

\bibitem{gordon2} J. M. Gordon, A. M. Goldman, and B. Whitehead, Phys.
Rev. Lett. \textbf{59}, 2311 (1987).

\bibitem{gordon3} J. M. Gordon and A. M. Goldman, Phys. Rev. B \textbf{35},
4909 (1987).

\bibitem{korshu} S. E. Korshunov, R. Meyer, and P. Martinoli, Phys. Rev. B 
\textbf{51}, 5914 (1995).

\bibitem{meyer} R. Meyer, S. E. Korshunov, Ch. Leemann, and P. Martinoli, 
Phys. Rev. B \textbf{66}, 104503 (2002).

\bibitem{nedellec} P. Nedellec, M. Aprili, J. Lesueur, and L. Dumoulin, 
Solid State Commun. \textbf{102}, 41 (1993).

\bibitem{arun2} A. Chakrabarti, Phys. Rev. B \textbf{60}, 10576 (1999).

\bibitem{arun3} A. Chakrabarti, Phys. Rev. B \textbf{72}, 134207 (2005).

\bibitem{schwalm4} W. Schwalm and B. J. Moritz, Phys. Rev. B \textbf{71},
134207 (2005).

\bibitem{arun1} A. Chakrabarti and B. Bhattacharyya, Phys. Rev. B
\textbf{56}, 13768 (1997).

\bibitem{kato} H. Kato and D. Yoshioka, Phys. Rev. B \textbf{50}, 4943
(1994).

\bibitem{kam} A. Kambili, C. J. Lambert, and J. H. Jefferson, Phys.
Rev. B \textbf{60}, 7684 (1999).

\bibitem{sant5} S. K. Maiti, Solid State Commun. \textbf{150}, 2212 (2010).

\bibitem{sant3} M. Dey, S. K. Maiti, and S. N. Karmakar, Org. Electron. 
\textbf{12}, 1017 (2011).

\end{thebibliography}
\end{document}